\documentclass[12pt,epsf]{article}
\usepackage{graphicx,amsmath,amssymb}
\begin{document}

\def\a{\alpha}
\def\b{\beta}
\def\c{\varepsilon}
\def\d{\delta}
\def\e{\epsilon}
\def\f{\phi}
\def\g{\gamma}
\def\h{\theta}
\def\k{\kappa}
\def\l{\lambda}
\def\m{\mu}
\def\n{\nu}
\def\p{\psi}
\def\q{\partial}
\def\r{\rho}
\def\s{\sigma}
\def\t{\tau}
\def\u{\upsilon}
\def\v{\varphi}
\def\w{\omega}
\def\x{\xi}
\def\y{\eta}
\def\z{\zeta}
\def\D{\Delta}
\def\G{\Gamma}
\def\H{\Theta}
\def\L{\Lambda}
\def\F{\phi}
\def\P{\Psi}
\def\S{\Sigma}

\def\o{\over}
\def\beq{\begin{eqnarray}}
\def\eeq{\end{eqnarray}}
\newcommand{\gsim}{ \mathop{}_{\textstyle \sim}^{\textstyle >} }
\newcommand{\lsim}{ \mathop{}_{\textstyle \sim}^{\textstyle <} }
\newcommand{\vev}[1]{ \left\langle {#1} \right\rangle }
\newcommand{\bra}[1]{ \langle {#1} | }
\newcommand{\ket}[1]{ | {#1} \rangle }
\newcommand{\EV}{ {\rm eV} }
\newcommand{\KEV}{ {\rm keV} }
\newcommand{\MEV}{ {\rm MeV} }
\newcommand{\GEV}{ {\rm GeV} }
\newcommand{\TEV}{ {\rm TeV} }
\def\diag{\mathop{\rm diag}\nolimits}
\def\Spin{\mathop{\rm Spin}}
\def\SO{\mathop{\rm SO}}
\def\O{\mathop{\rm O}}
\def\SU{\mathop{\rm SU}}
\def\U{\mathop{\rm U}}
\def\Sp{\mathop{\rm Sp}}
\def\SL{\mathop{\rm SL}}
\def\tr{\mathop{\rm tr}}

\def\IJMP{Int.~J.~Mod.~Phys. }
\def\MPL{Mod.~Phys.~Lett. }
\def\NP{Nucl.~Phys. }
\def\PL{Phys.~Lett. }
\def\PR{Phys.~Rev. }
\def\PRL{Phys.~Rev.~Lett. }
\def\PTP{Prog.~Theor.~Phys. }
\def\ZP{Z.~Phys. }

\newcommand{\bea}{\begin{eqnarray}}   
\newcommand{\eea}{\end{eqnarray}}
\newcommand{\bear}{\begin{array}}  
\newcommand {\eear}{\end{array}}
\newcommand{\bef}{\begin{figure}}  
\newcommand {\eef}{\end{figure}}
\newcommand{\bec}{\begin{center}}  
\newcommand {\eec}{\end{center}}
\newcommand{\non}{\nonumber}  
\newcommand {\eqn}[1]{\beq {#1}\eeq}
\newcommand{\la}{\left\langle}  
\newcommand{\ra}{\right\rangle}
\newcommand{\ds}{\displaystyle}
\def\SEC#1{Sec.~\ref{#1}}
\def\FIG#1{Fig.~\ref{#1}}
\def\EQ#1{Eq.~(\ref{#1})}
\def\EQS#1{Eqs.~(\ref{#1})}
\def\REF#1{(\ref{#1})}
\def\TEV#1{10^{#1}{\rm\,TeV}}
\def\GEV#1{10^{#1}{\rm\,GeV}}
\def\MEV#1{10^{#1}{\rm\,MeV}}
\def\KEV#1{10^{#1}{\rm\,keV}}
\def\lrf#1#2{ \left(\frac{#1}{#2}\right)}
\def\lrfp#1#2#3{ \left(\frac{#1}{#2} \right)^{#3}}


\baselineskip 0.7cm

\begin{titlepage}

\begin{flushright}
TU-888\\
IPMU11-0134\\
UT-11-29
\end{flushright}

\vskip 1.35cm
\begin{center}
{\large \bf 
Higgs mass and inflation
}
\vskip 1.2cm
Kazunori Nakayama$^a$
and
Fuminobu Takahashi$^{b,c}$

\vskip 0.4cm

{\it $^a$Department of Physics, University of Tokyo, Tokyo 113-0033, Japan}\\
{\it $^b$Department of Physics, Tohoku University, Sendai 980-8578, Japan}\\
{\it $^c$Institute for the Physics and Mathematics of the universe,
University of Tokyo, Kashiwa 277-8568, Japan}\\

\vskip 1.5cm

\abstract{ 
We show  that the standard-model Higgs boson mass $m_h$ is correlated with
the spectral index of density perturbation $n_s$ in the  inflation scenario 
with the inflaton being identified with the B$-$L Higgs boson. 
The Higgs boson mass ranges from $m_h \simeq 120$\,GeV to $140$\,GeV
for $n_s \simeq 0.95 - 0.96$. 
In particular, as $n_s$ approaches to $0.96$,
the Higgs mass is predicted to be in the range of $125$\,GeV to $140$\,GeV in the case of relatively light gauginos,
and $120$\,GeV to $135$\,GeV in the case where all SUSY particle masses are of the same order.
This will be tested soon by the LHC experiment and the Planck satellite. 
The relation is due to the PeV-scale supersymmetry
required by the inflationary dynamics. 
We also comment on the cosmological implications of our scenario such as non-thermal 
leptogenesis and dark matter.
}
\end{center}
\end{titlepage}

\setcounter{page}{2}

The inflationary paradigm~\cite{Guth:1980zm} has been well established
so far. However, despite its great success, 
 it is not fully known how the inflation occurred, how the Universe was reheated after inflation, 
how the dark matter as well as the baryon asymmetry were created.

If the inflaton is a gauge singlet field with very weak couplings to the standard-model
particles, it would be challenging to pin down the inflation model.
Instead, let us focus on a new inflation model recently proposed by the present authors~\cite{Nakayama:2011ri},
in which the B$-$L Higgs boson plays the role of the inflaton. 
The theoretical framework is  the minimal extension of the standard model (SM), namely, 
SM + right-handed neutrinos + gauged U(1)$_{\rm B-L}$. The small but non-zero neutrino masses
can be explained beautifully by the seesaw mechanism~\cite{seesaw}, if there are
  heavy right-handed neutrinos. 
With the addition of the three right-handed neutrinos, it is
reasonable to introduce the U(1)$_{\rm B-L}$ gauge symmetry, because it
is required by the charge quantization condition and is also motivated
by the GUT gauge group such as SO(10).  
Thus, we consider the framework, SM+$\nu_R$+U(1)$_{\rm B-L}$, as the minimal extension of
the SM. 
There are models where the GUT Higgs, including U(1)$_{\rm B-L}$ Higgs, is identified as
the waterfall field in hybrid inflation~\cite{hep-ph/9406319}.
In Ref.~\cite{Nakayama:2011ri} it has been shown that the inflation model using the B$-$L Higgs boson 
works successfully if there is  supersymmetry (SUSY) at a scale below $\sim 10^3$\,TeV $=$ PeV.
The presence of SUSY below the PeV scale 
is crucial for canceling the Coleman-Weinberg (CW) potential~\cite{Coleman:1973jx}
arising from the B$-$L gauge boson loop.\footnote{
  The possibility that the gauge non-singlet inflaton is protected
  from radiative corrections by SUSY was pointed out in
  Ref.~\cite{Ellis:1982ed}.}

Since it is conceivable that the inflationary dynamics has affected the selection of our Universe
in the string landscape, the SUSY breaking scale may also be  determined by the inflationary selection.\footnote{
Considering that weak-scale SUSY is not free of fine-tunings and  typically 
requires a fine-tuning at the percent level for the correct electroweak breaking
and that the observed cosmological constant lies in the anthropic window~\cite{Weinberg:1987dv},
we do not rely on the conventional naturalness argument in this letter.
}
Indeed, if there is a bias toward larger SUSY breaking scale, we expect that the upper bound
on the SUSY breaking scale is saturated. The precise value of the upper bound depends on the
detailed structure of the inflaton potential, and so, it is related with the properties of density perturbation
such as the spectral index.
Since the PeV-scale SUSY gives rise to sizable radiative corrections
to the Higgs mass, we can derive a non-trivial relation between the Higgs mass $m_h$
and the spectral index $n_s$. This is the main result of this letter.

\vspace{5mm}

Here we summarize the results of the inflation model in Ref.~\cite{Nakayama:2011ri}.
We present here the non-SUSY version for simplicity,
although it is possible to write down the model in a supersymmetric language~\footnote{
 The potential (\ref{Vinf}) with $m=2n$ can be derived from the superpotential,
 $W = X(v^2 - k (\phi {\bar \phi})^n)$,  with an appropriate K\"ahler potential,
 where $\phi$ and ${\bar \phi}$ denote a conjugate pair of superfields transforming under
 U(1)$_{\rm B-L}$. This form of the superpotential is ensured by assigning a discrete $Z_n$
 symmetry under which ${\bar \phi}$ is charged. Details are found in 
	Ref.~\cite{Nakayama:2011ri}. 
}.
 Let us consider an inflaton potential given by
\beq
V(\varphi)\;=\;V_0 - \frac{\kappa}{2n} \frac{\varphi^{2n}}{M_*^{2n-4}}
			+ \frac{\lambda}{2m} \frac{\varphi^{2m}}{M_*^{2m-4}},      \label{Vinf}
\label{infV}
\eeq
where $\kappa$ and $\lambda$ are numerical coefficients, 
$m$ and $n$ are integers satisfying $m > n \geq 2$ and $M_*$ is
a cut-off scale of the theory. We expect $M_*$ to be not far from the
GUT scale $\sim \GEV{15-16}$.
Here we have focused on the radial component of the B$-$L Higgs boson $\phi$,
\beq
\varphi \;\equiv\; \sqrt{2} |\phi|.
\eeq
In SUSY, the inflaton actually corresponds to the D-flat direction of U(1)$_{\rm B-L}$.
This potential has a global minimum at $\varphi=\varphi_{\rm min}$ given by
\beq
\varphi_{\rm min} \;=\; \lrfp{\kappa}{\lambda}{\frac{1}{2(m-n)}} M_*,     \label{phimin}
\eeq
which gives the U(1)$_{\rm B-L}$ symmetry breaking scale at low
energy.  We fix this scale to be $\GEV{15}$ as suggested by the atmospheric neutrino oscillations and
the seesaw mechanism~\cite{seesaw}.

The inflation takes place if the initial position of $\varphi$ is
sufficiently close to the origin, which is expected to be the case if there is thermal plasma
before the inflation. The scalar spectral index, $n_s$, is given by
\beq
1-n_s \;=\;  \frac{2}{1+N\frac{2n-2}{2n-1}},   \label{ns}
\eea
where $N$ is the e-folding number.
In the limit $n\gg 1$, it approaches to $n_s = 0.96$ for $N = 50$,
which is close to the center value of the WMAP
result~\cite{Komatsu:2010fb}. For $n = 3$, the spectral index is
about $0.95$, which is also consistent with observation. 

The reheating takes place through the coupling with the
right-handed neutrinos,
\beq
{\cal L} \;=\; -\frac{y_N}{2} \phi \,{\bar {\nu}_R^c} \nu_R + {\rm h.c.},
\label{phinur}
\eeq
which generates large Majorana masses for $\nu_R$. The coupling constant
$y_N $ is expected to be order unity for the heaviest $\nu_R$. 
The non-thermal leptogenesis~\cite{Asaka:1999yd} works successfully
for $n \geq 3$~\cite{Nakayama:2011ri}.  So we focus on the case of $n\geq 3$,
although we include the case of $n=2$ in our analysis for completeness.

So far we have used the inflaton potential at the tree-level.
The inflaton, the B$-$L Higgs boson,
necessarily couples to the U(1)$_{\rm B-L}$ gauge boson.
Furthermore, it is  coupled to the right-handed neutrinos to generate 
large Majorana masses. Due to these interactions,
 the inflaton potential receives corrections at the one-loop level. 
In fact, it is well known that the CW potential arising from the gauge
boson loop makes the effective potential so steep that the resultant
density perturbation becomes much larger than the observed
one~\cite{Starobinsky:1982ee}. One way to cancel or suppress the CW potential is to introduce SUSY.
The successful inflationary dynamics requires the following inequality to be satisfied~\cite{Nakayama:2011ri}:
\begin{equation}
	{\rm max}\left[\tilde{m}_\lambda, \tilde{m}_{N}\right] \; \lsim \;  0.1\,H_{\rm inf}.
	\label{bound}
\end{equation}
where $H_{\rm inf}$ is the Hubble scale during inflation,
$\tilde{m}_\lambda$ and $\tilde{m}_N$ represent the soft SUSY breaking masses of the 
B$-$L gaugino and the right-handed sneutrino, respectively. We have here approximated
the U(1)$_{\rm B-L}$ gauge coupling as well as the coupling of $\phi$ to the heaviest right-handed
neutrino to be of order unity. The effect of supergravity is also negligible if a similar 
inequality, $m_{3/2} \lesssim H_{\rm inf}$, is satisfied~\cite{Nakayama:2011ri}.

\begin{figure}[tbp]
\begin{center}
\includegraphics[scale=1.0]{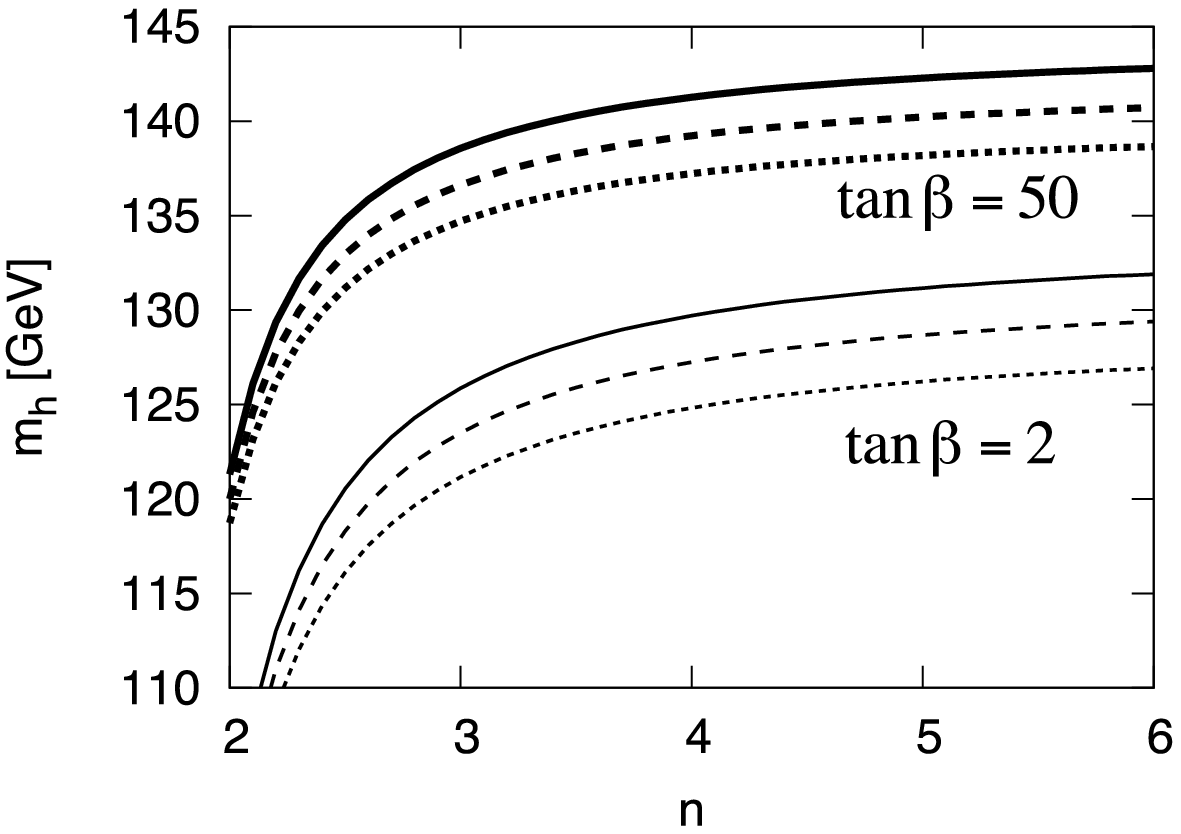}
\vskip 1cm
\includegraphics[scale=1.0]{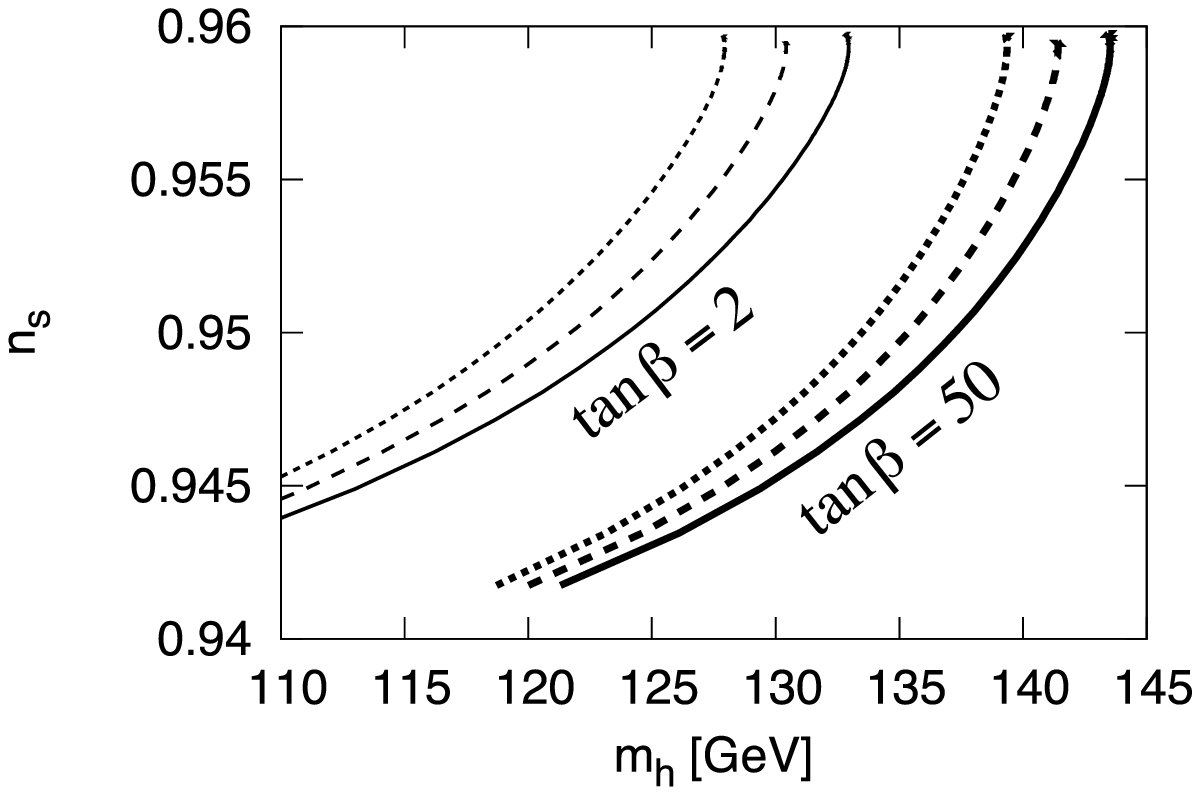}
\caption{
 (Top) The prediction for Higgs mass as a function of $n$
 for a different set of tan$\beta$ and $m_t$ in split SUSY scenario,
 where gaugino and higgsino masses are set to be 1\,${\rm TeV}$.
 Thick (thin) lines correspond to $\tan\beta=50$ $(2)$, and 
 solid, dashed, dotted lines correspond to $m_t = 175$, $173$, $171$\,GeV, respectively.
 We set $m=n+1$,
 $\varphi_{\rm min}= \GEV{15}$ and $M_*= \GEV{16}$. The results remain almost intact
if we change  the values of $m$ and $M_*$. 
(Bottom)
The correlation between the spectral index $n_s$ and Higgs mass. 
Meanings of each line are same as the top panel.
 }
\label{fig:mh}
\end{center}
\end{figure}

\begin{figure}[tbp]
\begin{center}
\includegraphics[scale=1.0]{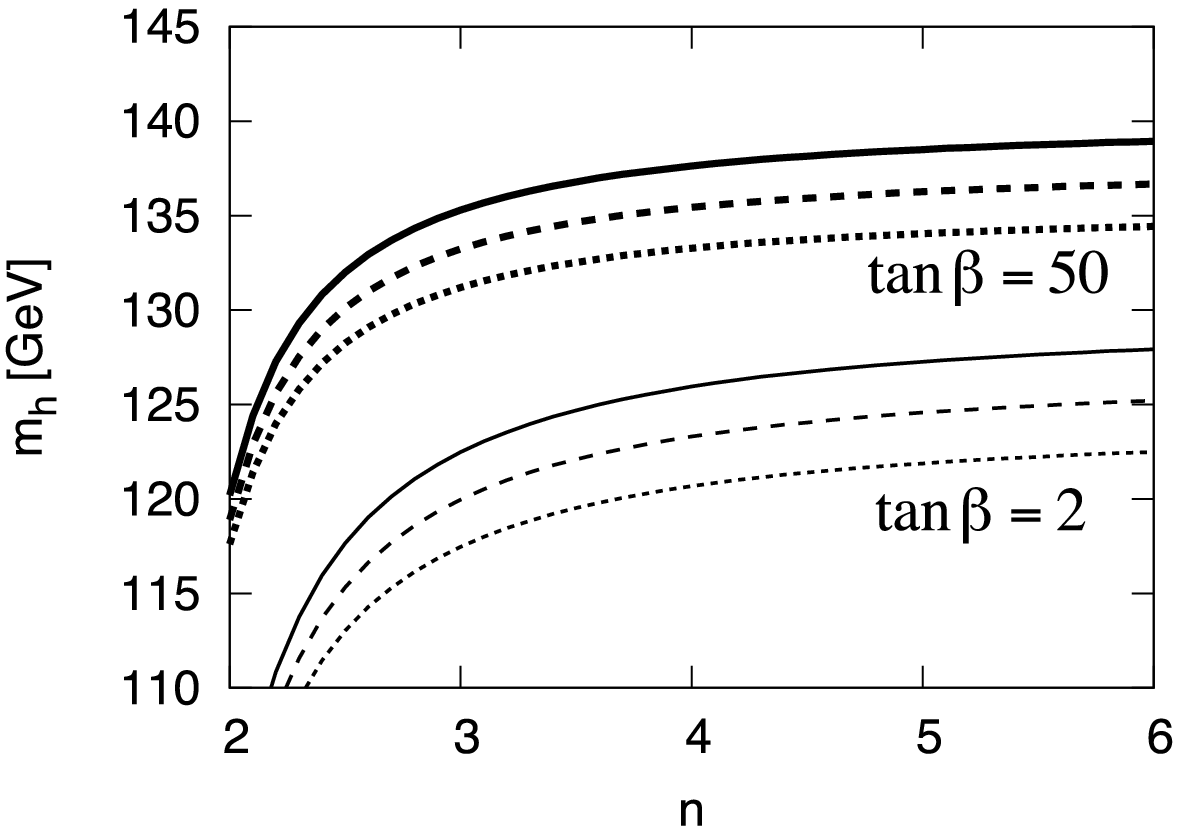}
\vskip 1cm
\includegraphics[scale=1.0]{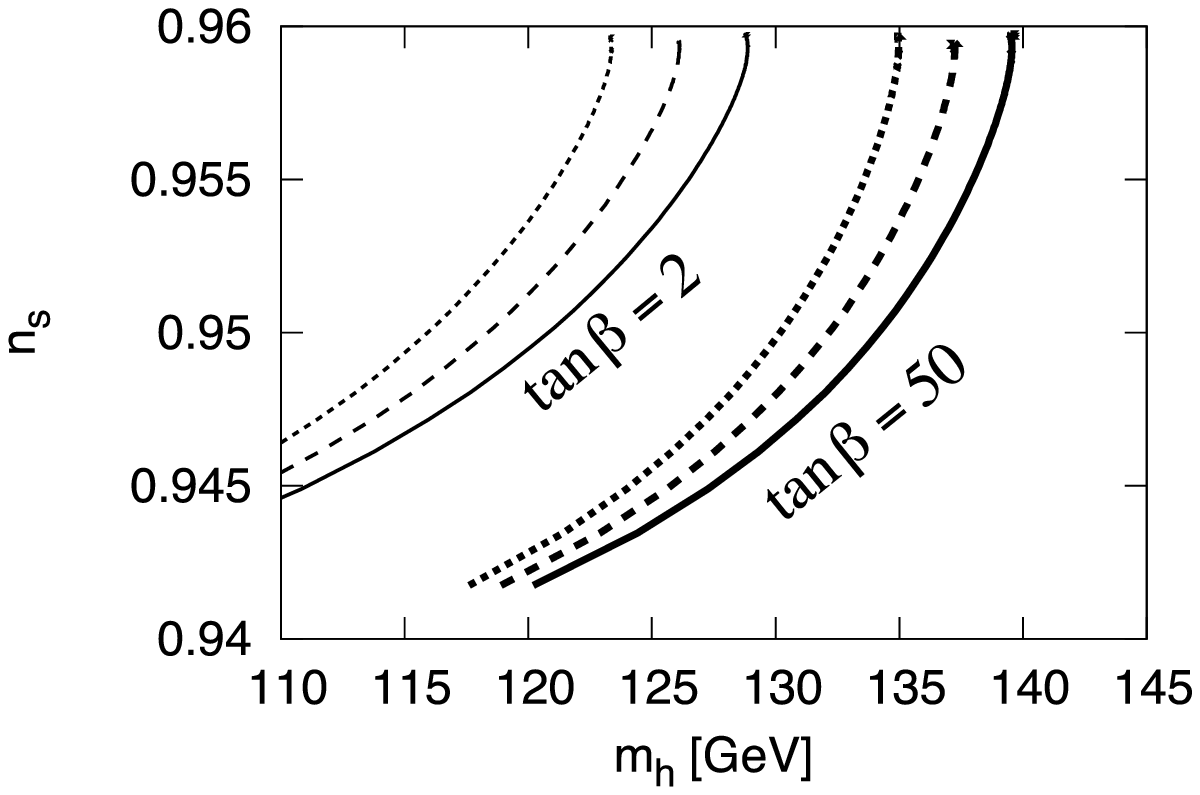}
\caption{
Same as Fig.~\ref{fig:mh}, but for the case where all SUSY particle mass scales are set to be $\tilde m$.
 }
\label{fig:mhSM}
\end{center}
\end{figure}

The inflationary scale is determined by the WMAP normalization on the density perturbation as 
$H_{\rm inf} \simeq 10^6 - 10^8$\,GeV depending on model parameters.
Assuming that the bound (\ref{bound}) is saturated, soft SUSY breaking masses for the SUSY SM (SSM) particles 
are expected to be much heavier than the weak scale.
This leads to relatively heavy lightest Higgs boson mass~\cite{ArkaniHamed:2004fb}.
We have calculated the lightest Higgs boson mass along the line of Ref.~\cite{Binger:2004nn}.
Fig.~\ref{fig:mh} shows the prediction for the Higgs mass as a function of $n$
for a different set of tan$\beta$ defined at the scale $\tilde m$. 
We assume that the bound \REF{bound} is saturated
and also that gauginos as well as higgsino masses are 1TeV while all other SUSY particles have masses of $\tilde m$.
The assumption on the gaugino and higgsino masses does not much affect the result as long as they are $\mathcal O(1)$\,TeV.
We set $m=n+1$, $\varphi_{\rm min}= \GEV{15}$ and $M_*= \GEV{16}$.
Thick (thin) lines correspond to $\tan\beta=50$ $(2)$, and 
 solid, dashed, dotted lines correspond to $m_t = 175$, $173$, $171$\,GeV, respectively.
In the bottom panel, we show the correlation between the spectral index $n_s$ and Higgs mass.
We note that the precise values of $m$ and $M_*$ are not relevant,
while changing the B$-$L breaking scale slightly affects the results. If we take $\varphi_{\rm min} = \GEV{14}$,
the predicted Higgs mass decreases by about $5$\,GeV at large $n$.
Focusing on $n\geq 3$, the Higgs mass ranges from $m_h = 120$\,GeV to $145$\,GeV, for $n_s = 0.95 - 0.96$. 
It is seen that the Higgs mass saturates at around $125-145$ GeV for large $n$.
Also the Higgs mass has a clear correlation with the scalar spectral index $n_s$.
Fig.~\ref{fig:mhSM} shows the same plots but for the case where all SUSY particle masses are set to be the scale of $\tilde m$.
In this case, the predicted Higgs mass is reduced :
$m_h = 115$\,GeV to $140$\,GeV, for $n_s = 0.95 - 0.96$.
This is mainly because the running of the Higgs quartic coupling does not receive correction from Higgs-higgsino-gaugino couplings
below the scale $\tilde m$.

\vspace{5mm}

The inflation model considered in this paper has interesting cosmological implications.
First, the inflaton mainly decays into the right-handed neutrinos through (\ref{phinur}).
Thus, the non-thermal leptogenesis~\cite{BA-90-78,Asaka:1999yd,hep-ph/0309134} works naturally. 
 The reheating temperature ranges from  $T_R \simeq \GEV{8}$ to $\GEV{10}$.
Secondly, the gravitino problem is avoided in our scenario. 
Assuming gravity or anomaly mediation, the gravitino mass $m_{3/2}$ is expected to be
about the PeV scale, and so, its effect on cosmology is very mild.
 As to the non-thermal gravitino production from the inflaton~\cite{Kawasaki:2006gs},
the branching ratio into the gravitinos are suppressed because the inflaton has a renormalizable
coupling with the right-handed neutrinos. 

In the case of split SUSY~\cite{ArkaniHamed:2004fb} with gauginos (especially  gluino) within the
reach of LHC, various types of signature may be observed~\cite{Alves:2011ug}.
In the anomaly mediation~\cite{Randall:1998uk}, the Wino is likely the lightest SUSY particle (LSP).
Assuming $m_{3/2} \simeq 1\,$PeV, the Wino mass is about $3$\,TeV. Interestingly,
the thermal relic of the Wino of mass $\sim 3$\,TeV can account for the observed 
dark matter abundance~\cite{Hisano:2006nn}. 
 This is an interesting coincidence: the SUSY breaking scale inferred from the inflation corresponds
to the one required by the thermal relic abundance of the Wino LSP.
In the gravity mediation, we expect that all the scalars and SSM gauginos have
comparable masses of PeV. Then, if the lightest SUSY particle (LSP) is the Bino-like neutralino, 
a small amount of the  R-parity violation may be needed  in order to avoid the overproduction of the LSP.
In this case the prime candidate for dark matter will be the QCD axion~\cite{Peccei:1977hh,Kim:1986ax}.

The inflation scenario we consider is the new inflation model~\cite{Linde:1981mu}, 
and its energy scale is so low that only a negligible amount of
the tensor mode is generated. Also non-Gaussianity of density perturbation as well as the running of the spectral
index is negligibly small.

In summary, in this letter we have shown that the standard-model Higgs boson mass and the spectral index of density perturbation, which are seemingly 
totally independent of each other, are actually related in the inflation scenario using the B$-$L Higgs boson as the inflaton.
This non-trivial relation will be soon checked by the LHC experiment and the Planck satellite.

\section*{Acknowledgment}

K.N. would like to thank M.~Ibe, T.~Moroi and T.~T.~Yanagida for important comments and discussion.
This work was supported by the Grant-in-Aid for Scientific Research on Innovative
Areas (No. 21111006) [KN and FT], Scientific Research (A)
(No. 22244030 [KN and FT] and No.21244033 [FT]), and JSPS Grant-in-Aid
for Young Scientists (B) (No. 21740160) [FT].  This work was also
supported by World Premier International Center Initiative (WPI
Program), MEXT, Japan.



\end{document}